\documentclass[twocolumn,aps,showpacs,floatfix,prc]{revtex4}
\usepackage{epsfig}
\begin{document}
\title{ Compact bifluid hybrid stars: Hadronic Matter mixed with self-interacting fermionic Asymmetric Dark Matter }

\author{ Somnath Mukhopadhyay$^{1*}$, Debasis Atta$^{2\S*}$, Kouser Imam$^{3\dagger*}$, D. N. Basu$^{4*}$ and C. Samanta$^{5**}$ }

\affiliation{$^*$Variable  Energy  Cyclotron  Centre, HBNI, 1/AF Bidhan Nagar, Kolkata 700 064, India }
\affiliation{$^{\S}$Govt. General Degree College, Kharagpur II, West Bengal 721 149, India }
\affiliation{$^\dagger$Dept. of Physics, Aliah University, IIA/27, New Town, Kolkata-700156, INDIA }
\affiliation{$^{**}$Dept. of Physics $\&$ Astronomy, Virginia Military Institute, Lexington, VA 24450, USA }

\email[E-mail 1: ]{somnathm@vecc.gov.in}
\email[E-mail 2: ]{debasisa906@gmail.com}
\email[E-mail 3: ]{kouserblackhole@gmail.com}
\email[E-mail 4: ]{dnb@vecc.gov.in}
\email[E-mail 5: ]{samantac@vmi.edu}
\date{\today }

\begin{abstract}

    The masses and radii of non-rotating and rotating configurations of pure hadronic stars mixed with self-interacting fermionic Asymmetric Dark Matter are calculated within the two-fluid formalism of stellar structure equations in general relativity. The Equation of State (EoS) of nuclear matter is obtained from the density dependent M3Y effective nucleon-nucleon interaction. We consider dark matter particle mass of 1 GeV. The EoS of self-interacting dark matter is taken from two-body repulsive interactions of the scale of strong interactions. We explore the conditions of equal and different rotational frequencies of nuclear matter and dark matter and find that the maximum mass of differentially rotating stars with self-interacting dark matter to be $\sim 1.94 M_\odot$ with radius $\sim 10.4$ kms. 


\end{abstract}

\maketitle

\noindent
\section{Introduction}
\label{section1}

    In the universe there are large empty regions and dense regions where the galaxies are distributed. This distribution is called the cosmic web that is speculated to be governed by the action of gravity on the invisible mysterious "dark matter". Recently, a research group led by Hiroshima University has suggested that the Cancer constellation has nine such large concentrations of dark matter, each the mass of a galaxy cluster \cite{Ut16}.
    
    Various theoretical models of dark matter are widespread, ranging from Cold Dark Matter to Warm Dark Matter to Hot Dark Matter and from Symmetric to Asymmetric Dark Matter \cite{Gi05,Ge84,Ki12,Ka14,Fr12}. Recent advances in cosmological precision tests further consolidate the minimal cosmological standard model, indicating that the universe contains 4.9$\%$ ordinary matter, 26.8$\%$ dark matter and 68.3$\%$ dark energy. Although being five times more abundant than ordinary matter, the basic properties of dark matter, such as particle mass and interactions are unsolved.

    A dark star composed mostly of normal matter and dark matter may have existed early in the universe before conventional stars were able to form. Those stars generate heat via annihilation reactions between the dark matter particles. This heat prevents such stars from collapsing into the relatively compact sizes of modern stars and therefore prevent nuclear fusion among the normal matter atoms from being initiated \cite{Do08}.
    
    One theory is that dark matter could be made of particles called axions. Unlike protons, neutrons and electrons that make up ordinary matter, axions can share the same quantum energy state. They also attract each other gravitationally, so they clump together. Dark matter is hard to study because it does not interact much with ordinary matter, but axion dark matter could theoretically be observed in the form of Bose stars \cite{Al15}. The Bose-Einstein condensation may come from the bosonic features of dark matter models. Phase transition to condensation can occur either when the temperature cools below critical value or when the density exceeds the critical value \cite{Li2012}. 

    The neutron stars could capture weakly interacting dark matter particles (WIMPs) because of their strong gravitational field, high density and finite, but very small, WIMP-to-nucleon cross section. In fact, if there is no baryon-dark matter interaction, purely baryonic neutron star would not capture dark matter at all. A dark star of comparable mass may as well accrete neutron star matter to form a dark matter dominated neutron star. In 1978, Steigman et al. \cite{St78} suggested that capture of WIMPs by individual stellar objects could affect the stellar structure and evolution. The effect of self annihilating dark matter on first-generation stars and on the evolution path of main sequence stars have been studied extensively \cite{Go89,Be08}. For non self-annihilating dark matter, its impact on main sequence stars \cite{Fr10} and neutron stars \cite{Ci11,Le11} have been studied in different dark matter models. Gravitational effects of non self-annihilating condensate dark matter on compact stellar objects has been studied \cite{LWC12} assuming dark matter as ideal Fermi gas and considering the accretion process through dark matter self-interaction from the surrounding halo. The non-annihilating heavy dark matter of mass greater than 1 GeV is predicted to accumulate at the center of neutron star leading it to a possible collapse \cite{La10}. The effect of this accumulation is observable only in cases where the annihilation cross section is extremely small \cite{Ko10,Ko11}. The capture is fully efficient even for WIMP-to-nucleon cross sections (elastic or inelastic) as low as 10$^{-18}$ mb. Moreover, a dark star of comparable mass may as well accrete neutron star matter to form a dark matter dominated neutron star. In addition to Axions and WIMPs, a general class of dark matter candidates called, Macros have been suggested that would have macroscopic size and mass \cite{Da15}.     
    
    Since dark matter interacts with normal baryonic matter through gravity, it is quite possible for white dwarfs and neutron stars to accrete dark matter and evolve to a dark matter admixed compact star \cite{Ko08,Be08,Mc10,Pe12,La10,Le11,Na06,Li12,Qi14}. The large baryonic density in compact stars increases the probability of dark matter capture within the star and eventually results in gravitational trapping.  It may also be possible for dark matter alone to form gravitationally bound compact objects and thus mimic stellar mass black holes \cite{Ha11}. 

    The hydrostatic equilibrium configuration of an admixture of degenerate dark matter and normal nuclear matter was studied by using a general relativistic two-fluid formalism taking non-self-annihilating dark matter particles of mass 1 GeV. A new class of compact stars was predicted that consisted a small normal matter core with radius of a few kilometers embedded in a ten-kilometer-sized dark matter halo \cite{Le11}.

    Compact objects formed by non-self annihilating dark matter admixed with ordinary matter has been predicted with Earth-like masses and radii from few kms to few hundred kms for weakly interacting dark matter. For the strongly interacting dark matter case, dark compact planets are suggested to form with Jupiter-like masses and radii of few hundred kms \cite{To15}. Possible implications of asymmetric fermionic dark matter for neutron stars has been studied that applies to various dark fermion models such as mirror matter models and to other models where the dark fermions have self-interactions \cite{Go13}.
    
    Although dark matter particles can have only very weak interactions with standard model states, it is an intriguing possibility that they experience much stronger self-interactions and thereby alter the behavior of dark matter on astrophysical and cosmological scales in striking ways. Recent studies \cite{Po15,Ho15,Ch16,Ja14,Ki14,Ro13} have provided constraints on dark matter self-interaction cross-section. The constraints are based on the Cusp-core problem and the ``Too big to fail" problem of galaxies. According to them the dark matter self-interaction cross-section per unit mass is about 0.1-100 cm$^2$/g  $\sim$0.1-1 barn/GeV, typical of the scale of strong interactions.
    
    In this work, we consider fermionic Asymmetric Dark Matter (ADM) particles of mass 1 GeV and the self-interaction mediator mass of 100 MeV (low mass implying strong interaction), mixed with rotating and non-rotating neutron stars. ADM, like ordinary baryonic matter, is charge asymmetric with only the dark baryon (or generally only the particle) excess remains after the annihilation of most antiparticles after the Big Bang. Hence these ADM particles are non self-annihilating and behaves like ordinary free particles. The gravitational stability and mass-radius relations of static, rigid and differentially rotating neutron stars mixed with fermionic ADM are calculated using the LORENE code \cite{Er10}. It is important to note that we do not allow any phase transition of the nuclear matter and the interaction between nuclear matter and dark matter is only through gravity.          

\noindent
\section{Equation of State of $\beta$-equilibrated nuclear matter}
\label{section2}

   The nuclear matter EoS is calculated using the isoscalar and the isovector \cite{La62,Sa83} components of M3Y interaction along with density dependence. The density dependence of this DDM3Y effective interaction is completely determined from nuclear matter calculations. The equilibrium density of the nuclear matter is determined by minimizing the energy per nucleon. The energy variation of the zero range potential is treated accurately by allowing it to vary freely with the kinetic energy part $\epsilon^{kin}$ of the energy per nucleon $\epsilon$ over the entire range of $\epsilon$. This is not only more plausible, but also yields excellent result for the incompressibility $K_\infty$ of the SNM which does not suffer from the superluminosity problem \cite{BCS08}. 
   
   In a Fermi gas model of interacting neutrons and protons, with isospin asymmetry $X= \frac{\rho_n - \rho_p} {\rho_n + \rho_p},~~~~\rho = \rho_n+\rho_p,$ where $\rho_n$, $\rho_p$ and $\rho$ are the neutron, proton and nucleonic densities respectively, the energy per nucleon for isospin asymmetric nuclear matter can be derived as \cite{BCS08}

\begin{equation}
 \epsilon(\rho,X) = [\frac{3\hbar^2k_F^2}{10m}] F(X) + (\frac{\rho J_v C}{2}) (1 - \beta\rho^n)  
\label{seqn1}
\end{equation}
\noindent
where $m$ is the nucleonic mass, $k_F$=$(1.5\pi^2\rho)^{\frac{1}{3}}$ which equals Fermi momentum in case of SNM, the kinetic energy per nucleon $\epsilon^{kin}$=$[\frac{3\hbar^2k_F^2}{10m}] F(X)$ with $F(X)$=$[\frac{(1+X)^{5/3} + (1-X)^{5/3}}{2}]$ and $J_v$=$J_{v00} + X^2 J_{v01}$, $J_{v00}$ and $J_{v01}$ represent the volume integrals of the isoscalar and the isovector parts of the M3Y interaction. The isoscalar $t_{00}^{M3Y}$ and the isovector $t_{01}^{M3Y}$ components of M3Y interaction potential are given by

\begin{eqnarray}
 t_{00}^{M3Y}(s, \epsilon)=&&+7999\frac{\exp( - 4s)}{4s}-2134\frac{\exp( -2.5s)}{2.5s} \nonumber \\
 && +J_{00}(1-\alpha\epsilon) \delta(s) \nonumber \\ 
 t_{01}^{M3Y}(s, \epsilon)=&&-4886\frac{\exp( - 4s)}{4s}+1176\frac{\exp( -2.5s)}{2.5s} \nonumber \\
 && +J_{01}(1-\alpha\epsilon) \delta(s)
\label{seqn2}
\end{eqnarray} 
\noindent
where $s$ represents the relative distance between two interacting nucleons, $J_{00}=-276$ MeV.fm$^3$, $J_{01}=+228$ MeV.fm$^3$ and the energy dependence parameter $\alpha=0.005$ MeV$^{-1}$. The strengths of the Yukawas were extracted by fitting its matrix elements in an oscillator basis to those elements of G-matrix obtained with Reid-Elliott soft core NN interaction and the ranges were selected to ensure OPEP tails in the relevant channels as well as a short-range part which simulates the $\sigma$-exchange process \cite{Be77}. The density dependence is employed to account for the Pauli blocking effects and the higher order exchange effects \cite{Sa79}. Thus the DDM3Y effective NN interaction is given by $v_{0i}(s,\rho, \epsilon) = t_{0i}^{M3Y}(s, \epsilon) g(\rho)$ where the density dependence $g(\rho) = C (1 - \beta \rho^n)$ \cite{BCS08} with $C$ and $\beta$ being the constants of density dependence.

    The Eq.(1) can be differentiated with respect to $\rho$ to yield equation for $X=0$:  
    
\begin{equation}
 \frac{\partial\epsilon}{\partial\rho} = [\frac{\hbar^2k_F^2}{5m\rho}] + \frac{J_{v00} C}{2} [1 - (n+1)\beta\rho^n] 
-\alpha J_{00} C [1 - \beta\rho^n]  [\frac{\hbar^2k_F^2}{10m}].
\label{seqn3}
\end{equation}
\noindent
The equilibrium density of the cold SNM is determined from the saturation condition. Then Eq.(1) and Eq.(3) with the saturation condition $\frac{\partial\epsilon}{\partial\rho} = 0$ at $\rho = \rho_{0}$, $\epsilon = \epsilon_{0}$ can be solved simultaneously for fixed values of the saturation energy per nucleon $\epsilon_0$ and the saturation density $\rho_{0}$ of the cold SNM to obtain the values of $\beta$ and $C$. The constants of density dependence $\beta$ and $C$, thus obtained, are given by 

\begin{equation}
 \beta = \frac{[(1-p)+(q-\frac{3q}{p})]\rho_{0}^{-n}}{[(3n+1)-(n+1)p+(q-\frac{3q}{p})]}
\label{seqn4}
\end{equation} 
\noindent
where $p$=$\frac{[10m\epsilon_0]}{[\hbar^2k_{F_0}^2]}$, $q$=$\frac{2\alpha\epsilon_0J_{00}}{J^0_{v00}}$, $J^0_{v00}$=$J_{v00}(\epsilon^{kin}_0)$ implying $J_{v00}$ at $\epsilon^{kin}$=$\epsilon^{kin}_0$, the kinetic energy part of the saturation energy per nucleon of SNM,  $k_{F_0}$=$[1.5\pi^2\rho_0]^{1/3}$ and 

\begin{equation}
 C = -\frac{[2\hbar^2k_{F_0}^2] }{ 5mJ^0_{v00} \rho_0 [1 - (n+1)\beta\rho_0^n -\frac{q\hbar^2k_{F_0}^2 (1-\beta\rho_0^n)}{10m\epsilon_0}]},
\label{seqn5}
\end{equation} 
\noindent
respectively. It is quite obvious that the constants of density dependence $C$ and $\beta$ obtained by this method depend on the saturation energy per nucleon $\epsilon_0$, the saturation density $\rho_{0}$, the index $n$ of the density dependent part and on the strengths of the M3Y interaction through the volume integral $J^0_{v00}$. 
        
    The calculations are performed using the values of the saturation density $\rho_0$=0.1533 fm$^{-3}$ \cite{Sa89} and the saturation energy per nucleon $\epsilon_0$=-15.26 MeV \cite{CB06} for the SNM obtained from the co-efficient of the volume term of Bethe-Weizs\"acker mass formula which is evaluated by fitting the recent experimental and estimated atomic mass excesses from Audi-Wapstra-Thibault atomic mass table \cite{Au03} by minimizing the mean square deviation incorporating correction for the electronic binding energy \cite{Lu03}. In a similar recent work, including surface symmetry energy term, Wigner term, shell correction and proton form factor correction to Coulomb energy also, $a_v$ turns out to be 15.4496 MeV and 14.8497 MeV when $A^0$ and $A^{1/3}$ terms are also included \cite{Ro06}. Using the usual values of $\alpha$=0.005 MeV$^{-1}$ for the parameter of energy dependence of the zero range potential and $n$=2/3, the values obtained for the constants of density dependence $C$ and $\beta$ and the SNM incompressibility $K_\infty$ are 2.2497, 1.5934 fm$^2$ and 274.7 MeV, respectively. The saturation energy per nucleon is the volume energy coefficient and the value of -15.26$\pm$0.52 MeV covers, more or less, the entire range of values obtained for $a_v$ for which now the values of $C$=2.2497$\pm$0.0420, $\beta$=1.5934$\pm$0.0085 fm$^2$ and the SNM incompressibility $K_\infty$=274.7$\pm$7.4 MeV.  
    
    The symmetric nuclear matter incompressibility $K_\infty$, nuclear symmetry energy at saturation density $E_{sym}(\rho_0)$, the slope $L$ and  isospin dependent part $K_\tau$ of the isobaric incompressibility are also tabulated in Table-I since these are all in excellent agreement with the recently extracted constraints from the measured isotopic dependence of the giant monopole resonances in even-A Sn isotopes \cite{Li07}, from the neutron skin thickness of nuclei, and from analyses of experimental data on isospin diffusion and isotopic scaling in intermediate energy heavy-ion collisions. 
    
    The calculations for masses and radii are performed using the EoS covering the crustal region of a compact star which are Feynman-Metropolis-Teller (FMT) \cite{FMT49}, Baym-Pethick-Sutherland (BPS) \cite{BPS71} and Baym-Bethe-Pethick (BBP) \cite{BBP71} upto number density of 0.0582 fm$^{-3}$ and $\beta$-equilibrated neutron star matter beyond. Figs.-1 and 2 represent the Mass-Central density and Mass-Radius plots respectively for slowly rotating pure neutron stars with DDM3Y EoS. The maximum mass goes to $1.9227 M_{\odot}$ with a radius of 9.7559 kms \cite{At17,At14,Se14}. 

\begin{table}[htbp]
\centering
\caption{Results of present calculations for $n$=$\frac{2}{3}$ of symmetric nuclear matter incompressibility $K_\infty$, nuclear symmetry energy at saturation density $E_{sym}(\rho_0)$, the slope $L$ and  isospin dependent part $K_\tau$ of the isobaric incompressibility (all in MeV) \cite{Ch09,Ba09} are tabulated.}
\begin{tabular}{cccc}
\hline
\hline
$K_\infty$&$E_{sym}(\rho_0)$&$L$&$K_\tau$ \\ 
\hline
 $274.7\pm7.4$&$30.71\pm0.26$&$45.11\pm0.02$&$-408.97\pm3.01$ \\
\hline
\hline
\end{tabular} 
\label{table1}
\end{table}
\noindent 

\begin{figure}[t]
\vspace{0.0cm}
\eject\centerline{\epsfig{file=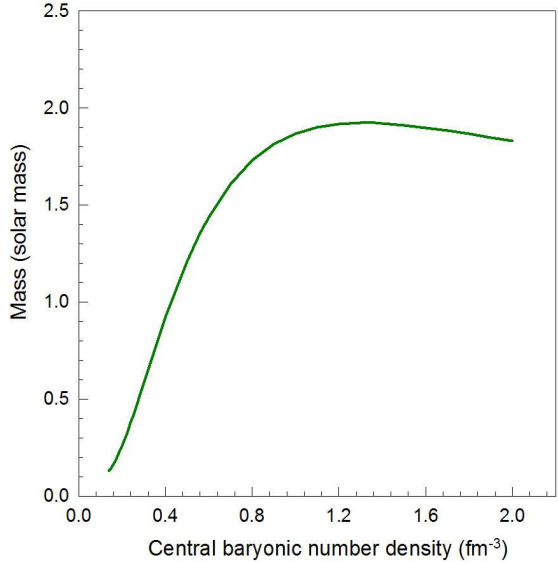,height=8cm,width=8cm}}
\caption
{Mass vs. central baryonic density plot of slowly rotating neutron stars for the DDM3Y EoS. } 
\label{fig1}
\vspace{0.0cm}
\end{figure}
\noindent

\begin{figure}[t]
\vspace{0.0cm}
\eject\centerline{\epsfig{file=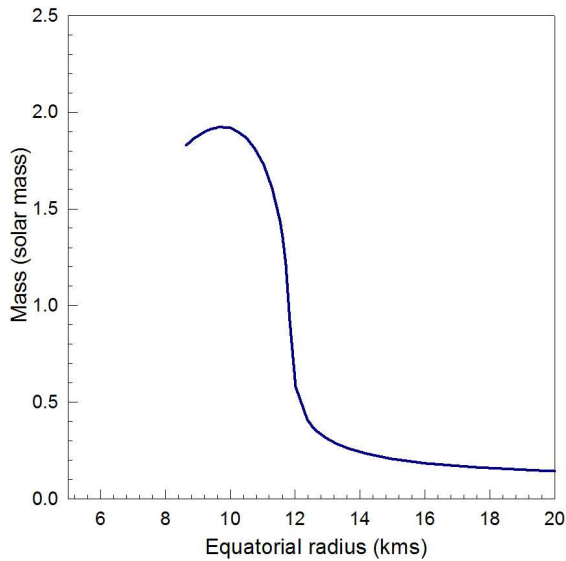,height=8cm,width=8cm}}
\caption
{Mass-equatorial radius plot of slowly rotating neutron stars for the DDM3Y EoS. } 
\label{fig2}
\vspace{0.0cm}
\end{figure}
\noindent     
    
\noindent
\section{ Equation of State of non-interacting fermionic Asymmetric Dark Matter }
\label{section3}
    
    We consider the non-interacting fermionic ADM to be a completely degenerate free Fermi gas of particle mass $m_\chi$ at zero temperature. By the Pauli exclusion principle, no quantum state can be occupied by more than one fermion with an identical set of quantum numbers. Thus a non-interacting Fermi gas, unlike a Bose gas, is prohibited from condensing into a Bose-Einstein condensate. The total energy of the Fermi gas at absolute zero is larger than the sum of the single-particle ground states because the Pauli principle implies a degeneracy pressure that keeps fermions separated and moving. 
    
    The non-interacting assembly of fermions at zero temperature exerts pressure because of kinetic energy from different states filled up to Fermi level. Since pressure is force per unit area which means rate of momentum transfer per unit area, it is given by    
    
\begin{equation}
 P_\chi= \frac{1}{3}\int p v n_p d^3p = \frac{1}{3}\int \frac{p^2c^2}{\sqrt{(p^2c^2+m_\chi^2c^4)}} n_p d^3p
\label{seqn6}
\end{equation}
where $m_\chi$ is the rest mass of dark particles, $v$ is the velocity of the particles with momentum $\vec p$ and $n_p d^3p$ is the number of particles per unit volume having momenta between $\vec p$ and $\vec p + d\vec p$. The factor $\frac{1}{3}$ accounts for the fact that, on average, only $\frac{1}{3}$rd of total particles $n_p d^3p$ are moving in a particular direction. For fermions having spin $\frac{1}{2}$, degeneracy = 2, $n_p d^3p = \frac{8\pi p^2 dp}{h^3}$ and hence number density $n_\chi$is given by

\begin{equation}
 n_\chi= \int_0^{p_F} n_p d^3p = \frac{8\pi p_F^3}{3h^3} = \frac{x_F^3}{3\pi^2\lambda_\chi^3}
\label{seqn7}
\end{equation}
where $p_F$ is the Fermi momentum which is maximum momentum possible at zero temperature, $x_F=\frac{p_F}{m_\chi c}$ is a dimensionless quantity and $\lambda_\chi=\frac{\hbar}{m_\chi c}$ is the Compton wavelength. The energy density $\varepsilon_\chi$ is given by

\begin{equation}
 \varepsilon_\chi= \int_0^{p_F} E n_p d^3p = \int_0^{p_F} \sqrt{(p^2c^2+m_\chi^2c^4)} \frac{8\pi p^2 dp}{h^3}
\label{seqn8}
\end{equation}
which, along with Eq.(7), turns out upon integration to be

\begin{equation}
 \varepsilon_\chi= \frac{m_\chi c^2}{\lambda_\chi^3} \chi(x_F); ~~~~  P_\chi= \frac{m_\chi c^2}{\lambda_\chi^3} \phi(x_F),
\label{seqn9}
\end{equation}
where

\begin{equation}
 \chi(x)= \frac{1}{8\pi^2}[x\sqrt{1+x^2}(1+2x^2)-\ln(x+\sqrt{1+x^2})]
\label{seqn10}
\end{equation}
and

\begin{equation}
 \phi(x)= \frac{1}{8\pi^2}[x\sqrt{1+x^2}(\frac{2x^2}{3}-1)+\ln(x+\sqrt{1+x^2})].
\label{seqn11}
\end{equation}
\noindent

\noindent
\section{ Equation of State of strongly self-interacting fermionic Asymmetric Dark Matter }
\label{section4}

    In order to calculate EoS of strongly interacting fermionic ADM we take course to massive vector field theory similar to the meson exchange of the nuclear interaction. The Lagrangian density (in natural units) of a massive vector field is given by
    
\begin{equation}
\mathcal{L}=-\frac{1}{4}F_{\mu\nu}F^{\mu\nu}+\frac{1}{2}m_I^2A_{\mu}A^{\mu}-j_{\mu}A^{\mu}
\label{seqn12}
\end{equation}
\noindent
where $F_{\mu\nu}=\partial_{\mu}A_{\nu}-\partial_{\nu}A_{\mu}$ , $A^{\mu}$ is the 4-vector field, $j^{\mu}$ is the 4-current and $m_I$ is the mass of the field quanta. The equation of motion is given by
 
\begin{equation}
(\partial_\nu\partial^\nu+m_I^2)A^{\mu}=j^{\mu}.
\label{seqn13}
\end{equation}
\noindent
Now considering a charge of magnitude $g$ at rest at the origin we have 
    
\begin{equation}
j^0=g\delta^3(\vec{x})  ~~~~ \vec{j}=0.
\label{seqn14}
\end{equation}
\noindent
Substituting the above in the right side of Eqn. (14) and also noting that $A^0=V$ and $\vec{A}=0$
we finally get 

\begin{equation}
(\nabla^2-m_I^2)V=-g\delta^3(\vec{x})
\label{seqn15}
\end{equation}   
\noindent
whose solution is the Yukawa potential:

\begin{equation}
V(r)=g\frac{e^{-m_Ir}}{4\pi r}
\label{seqn16}
\end{equation}   
\noindent 
Hence the potential energy of two like charges of magnitude $g$ is 

\begin{equation}
V_{12}(r)=g^2\frac{e^{-m_Ir}}{4\pi r}
\label{seqn17}
\end{equation}
\noindent 
and is repulsive in nature.

    To proceed to the EoS, we calculate the total energy of a system of particles classically by summing over the interactions of all pairs of particles. To facilitate the calculation, we assume that the macroscopic assembly is uniformly distributed, thereby neglecting the influence of the interaction on the mean inter-particle separation. In other words, we ignore any correlations between particle positions due to their mutual interaction. Finally, we assume that the number of particles is sufficiently large that we can replace sums by integrals, and that the characteristic size of the assembly $R$ satisfies $R >> 1/m_I$ \cite{St}.
      
    The total Yukawa potential energy of a system of $N$ particles in volume $\Omega$ is
    
\begin{equation}
E_{\Omega}=\frac{1}{2}\sum_{i\neq j}V_{ij}=\frac{1}{2}n^2g^2\int\int\frac{e^{-m_Ir_{ij}}}{4\pi r_{ij}}d\Omega_id\Omega_j,
\label{seqn18}
\end{equation}
\noindent
where $n$ is the number density.
   
    Choosing one particle at the origin and integrating to infinity (ignoring surface terms) we get,
    
\begin{equation}
E_{\Omega}=\frac{1}{2m_I^2}n^2g^2\Omega,
\label{seqn19}
\end{equation}          
\noindent
so that the interaction energy density can be written as,

\begin{equation}
\varepsilon_{int}=\frac{E_{\Omega}}{\Omega}=\frac{1}{2m_I^2}n^2g^2.
\label{seqn20}
\end{equation}
\noindent
Now putting $g^2/2=1$ for convenience, $x_f=k_f/m_\chi$, where $m_\chi$ is the rest mass of the dark matter particle and using the relation $k_f=(3\pi^2n)^{1/3}$ we get putting back $\hbar$ and $c$

\begin{equation}
\varepsilon_{int}=\left(\frac{1}{3\pi^2}\right)^2\frac{x_f^6m_\chi^6}{(\hbar c)^3m_I^2}
\label{seqn21}
\end{equation} 
\noindent
where $m_\chi$ and $m_I$ are expressed in MeV.

    The pressure due to the interacting energy density can be computed with the help of the thermodynamic relation $P_{int}=n^2\frac{d}{dn}\Big(\frac{\varepsilon_{int}}{n}\Big)$, which yields
    
\begin{equation}
P_{int}=\left(\frac{1}{3\pi^2}\right)^2\frac{x_f^6m_\chi^6}{(\hbar c)^3m_I^2}
\label{seqn22}
\end{equation}
\noindent

    Hence the total energy density and pressure of self-interacting dark matter particles are given by
    
\begin{equation}
\varepsilon_{\chi int}=\varepsilon_\chi+ \varepsilon_{int}=\frac{m_\chi}{\lambda_\chi^3} \chi(x_F)+\left(\frac{1}{3\pi^2}\right)^2\frac{x_f^6m_\chi^6}{(\hbar c)^3m_I^2}
\label{seqn23}
\end{equation}
\noindent
 
\begin{equation} 
 P_{\chi int}=P_\chi+ P_{int}= \frac{m_\chi}{\lambda_\chi^3} \phi(x_F)+\left(\frac{1}{3\pi^2}\right)^2\frac{x_f^6m_\chi^6}{(\hbar c)^3m_I^2} 
\label{seqn24}
\end{equation}
\noindent

    The mass of the exchange boson determines the strength and range of the interaction implying lower the mass stronger the interaction and for non-interacting dark matter, $m_I$ is infinity and second terms in above equations are absent.
		
		Figs.-3 and 4 depict the plots of mass vs. central dark matter density and mass vs. equatorial radius respectively for static and rotating stars using self-interacting dark matter EoS. We see that the maximum mass for non-rotating stars goes to $3.0279 M_{\odot}$ with a radius of 16.2349 kms and that for rotating stars goes to $3.1460 M_{\odot}$ with equatorial radius of 19.2173 kms. Now, if we take the dark matter particle mass $m_\chi$ to be 0.5 GeV, then the maximum mass goes to $\sim 12.6 M_{\odot}$ using the relation Mass $\propto 1/m^2_\chi$ \cite{Na06}, thus mimicking stellar mass black holes.  		
		
\begin{figure}[t]
\vspace{0.0cm}
\eject\centerline{\epsfig{file=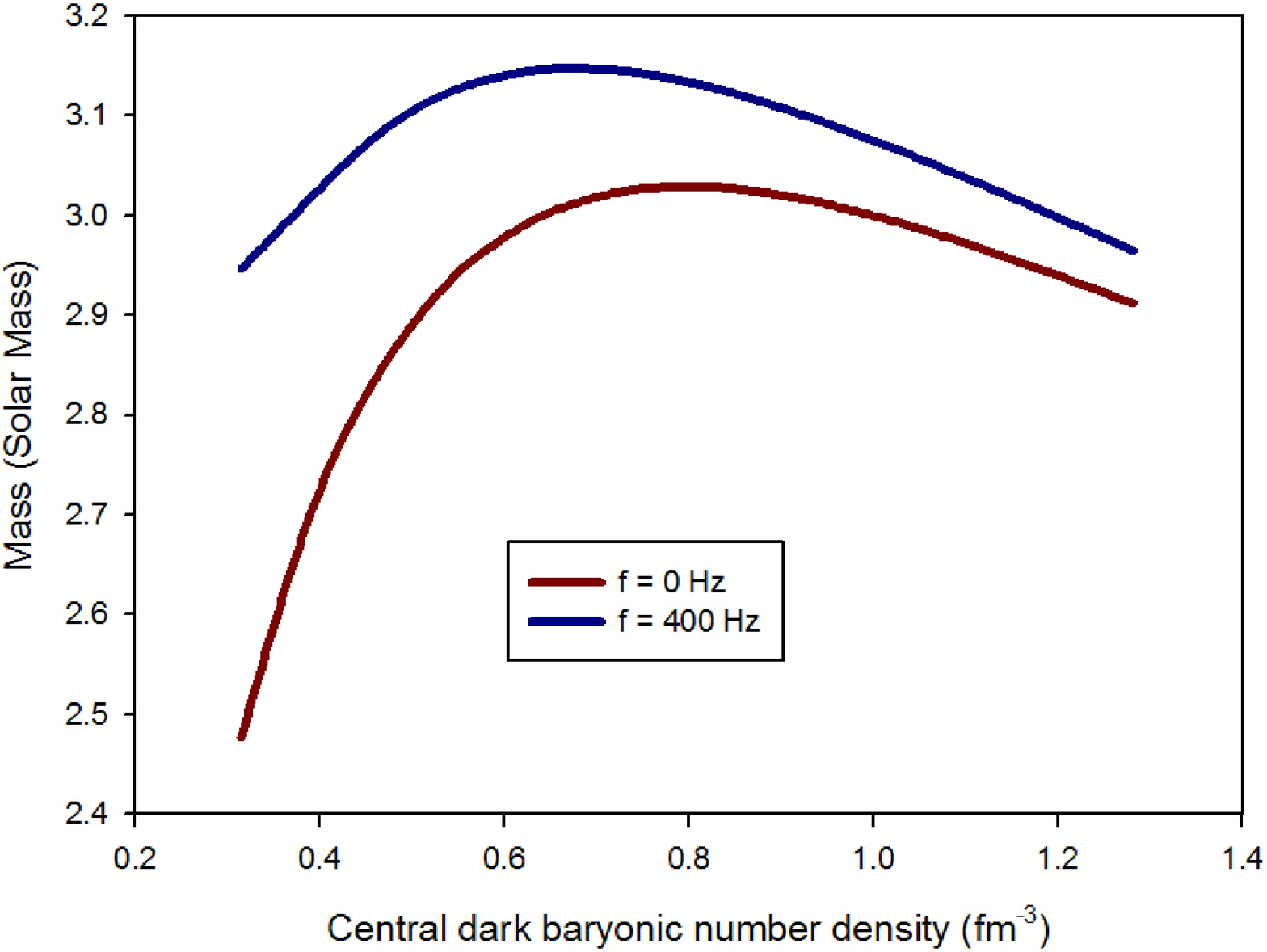,height=8cm,width=8cm}}
\caption
{Plots of mass vs. central density for static and rotating fermionic Asymmetric Dark Matter stars. } 
\label{fig3}
\vspace{0.0cm}
\end{figure}
\noindent

\begin{figure}[t]
\vspace{0.0cm}
\eject\centerline{\epsfig{file=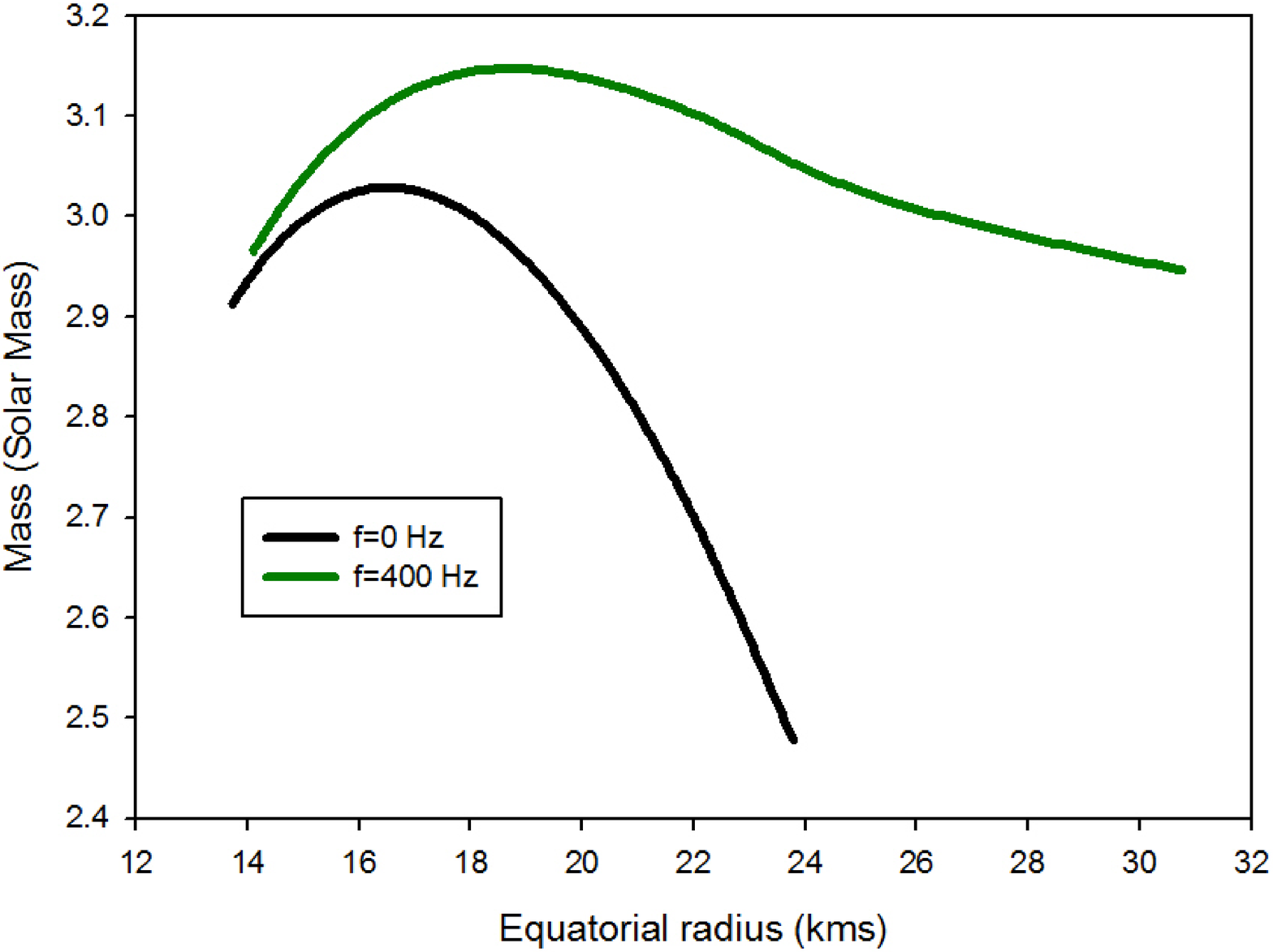,height=8cm,width=8cm}}
\caption
{Mass-equatorial radius plots for static and rotating fermionic Asymmetric Dark Matter stars. } 
\label{fig4}
\vspace{0.0cm}
\end{figure}
\noindent 

\noindent    
\section{ Two-fluid TOV equation }
\label{section5}

    We consider two ideal fluids - the nuclear matter and fermionic dark matter with the above two EoSs coupled gravitationally to form the structure of the mixed neutron star. The energy-momentum tensor of the mixed fluid can be written as \cite{Sa09,Go13}
    
\begin{eqnarray}
T^{\mu \nu}= &&T^{\mu \nu}_{nuc}+ T^{\mu \nu}_{dark}=(\varepsilon_{nuc}+P_{nuc})u_1^{\mu}u_1^{\nu}-P_{nuc}g^{\mu \nu} \nonumber\\
+&&(\varepsilon_{dark}+P_{dark})u_2^{\mu}u_2^{\nu}-P_{dark}g^{\mu \nu}
\label{seqn25}
\end{eqnarray}    
\noindent 
where $u_1^{\mu}$, $\varepsilon_{nuc}$ and $P_{nuc}$ are the 4-velocity, energy density and pressure of nuclear matter respectively while the corresponding quantities in the second term are of dark matter.

    For non-rotating case the metric is spherically symmetric and the hydrostatic equations of the two fluids can be written as coupled two-fluid Tolman-Oppenheimer-Volkoff (TOV) equations

\begin{eqnarray}
\frac{dP_{nuc}(r)}{dr}=-\frac{GM(r)\rho_{nuc}(r)}{r^2}\left(1+\frac{P_{nuc}}{\varepsilon_{nuc}}\right)\times \nonumber\\ 
\left(1+\frac{4\pi r^3(P_{nuc}+P_{dark})}{M(r)c^2}\right)\left(1-\frac{2GM(r)}{rc^2}\right)^{-1}  \nonumber\\
\frac{dP_{dark}(r)}{dr}=-\frac{GM(r)\rho_{dark}(r)}{r^2}\left(1+\frac{P_{dark}}{\varepsilon_{dark}}\right)\times\nonumber\\ 
\left(1+\frac{4\pi r^3(P_{nuc}+P_{dark})}{M(r)c^2}\right)\left(1-\frac{2GM(r)}{rc^2}\right)^{-1}  \nonumber\\
\frac{dM_{nuc}(r)}{dr}=4\pi r^2\rho_{nuc}(r) \nonumber\\
\frac{dM_{dark}(r)}{dr}=4\pi r^2\rho_{dark}(r)  \nonumber\\
M(r)=M_{nuc}(r)+M_{dark}(r)
\label{seqn26}
\end{eqnarray} 
\noindent         
where $\rho_{nuc}=\varepsilon_{nuc}/c^2$, $M_{nuc}$ is the mass density and total mass of nuclear matter while the corresponding quantities in the second equation are of dark matter. $M(r)$ is the total mass of nuclear and dark matter.

\noindent
\section{ Theoretical calculations }
\label{section6}

    The mass-radius relationship of non-rotating, rigidly rotating and differentially rotating neutron stars admixed with dark matter is calculated using the LORENE code. The nuclear matter and dark matter EoSs are fitted to a polytropic form $P=K\rho^{\gamma}$ where $P$ is the pressure, $\rho$ is the mass density, $K$ the polytropic constant and $\gamma$ the polytropic index for the corresponding fluid. For interacting nuclear matter $\gamma=2.03$ and $K=5.65283\times10^{35}$ in C.G.S. units. For interacting dark matter $\gamma=1.97562$ and $K=1.33404\times10^{36}$ in C.G.S. units. We take dark matter particle mass to be of 1 GeV and the exchange boson mass $m_I=100$ MeV, typical of strong interaction. First, we keep the dark matter central enthalpy to be $0.24c^2$ (fixed) and vary the nuclear matter central enthalpy for static, rigidly rotating and differentially rotating configurations and next we reverse the roles of nuclear and dark matter.   
    
\begin{figure}[t]
\vspace{0.0cm}
\eject\centerline{\epsfig{file=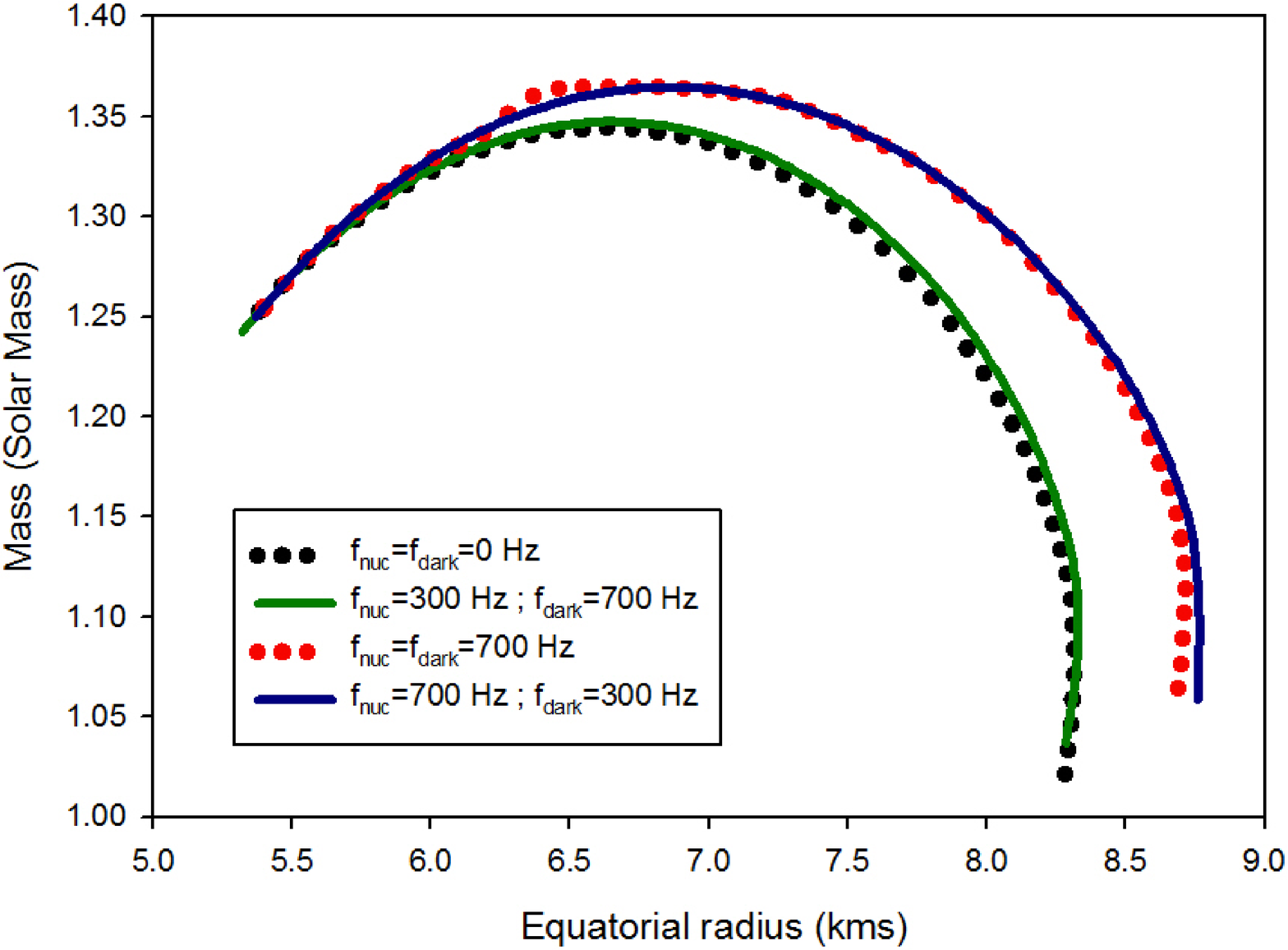,height=8cm,width=8cm}}
\caption
{Plots of total mass vs. equatorial radius of static, rigidly rotating and differentially rotating neutron stars mixed with interacting fermionic Asymmetric Dark Matter with fixed dark matter central enthalpy ($0.24c^2$) and varying nuclear matter central enthalpies. } 
\label{fig5}
\vspace{0.0cm}
\end{figure}
\noindent 

\begin{figure}[t]
\vspace{0.0cm}
\eject\centerline{\epsfig{file=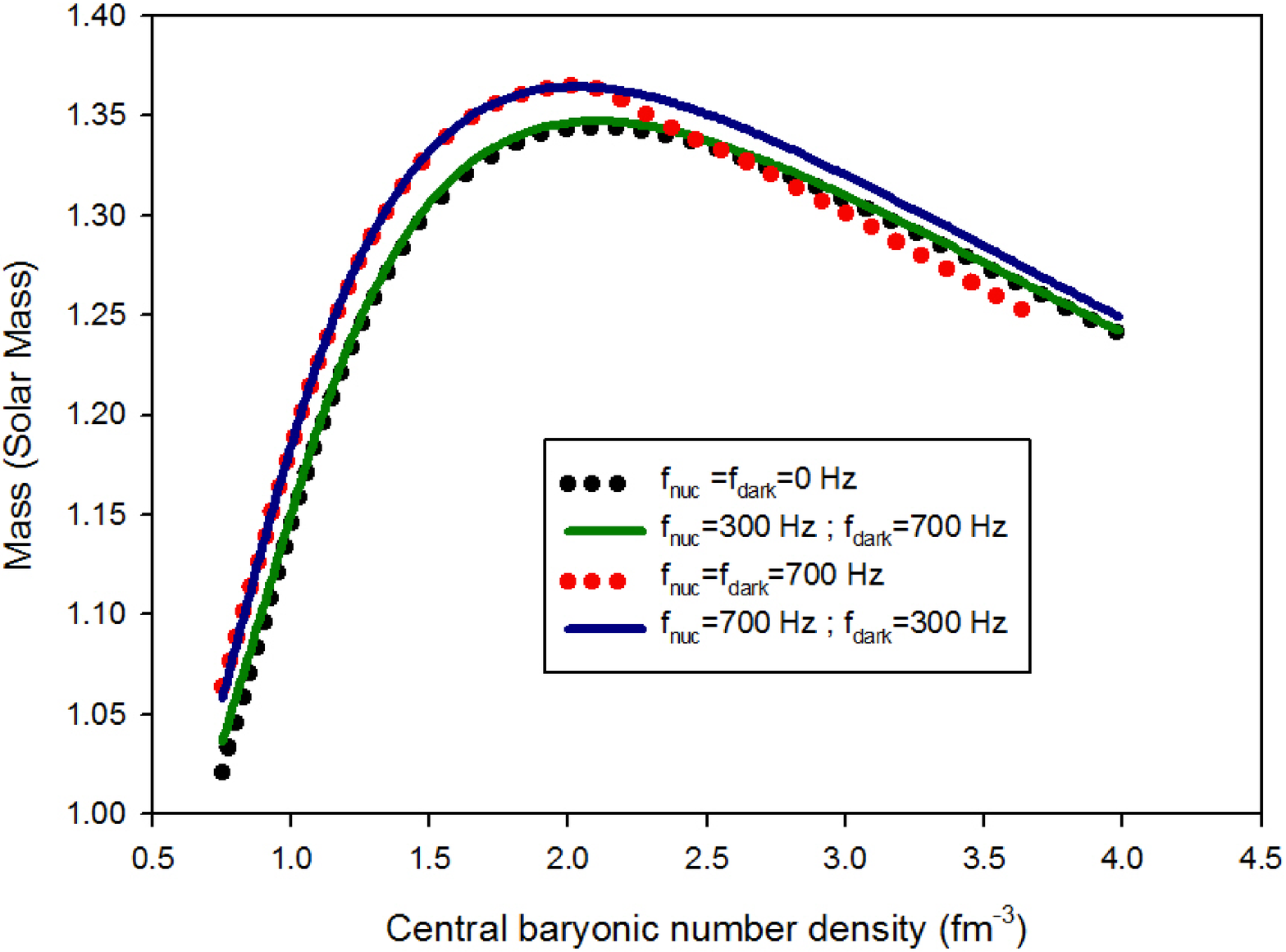,height=8cm,width=8cm}}
\caption
{Plots of total mass vs. central baryonic density of static, rigidly rotating and differentially rotating neutron stars mixed with self-interacting fermionic Asymmetric Dark Matter with fixed dark matter central enthalpy ($0.24c^2$) and varying nuclear matter central enthalpies.}
\label{fig6}
\vspace{0.0cm}
\end{figure}
\noindent

\begin{figure}[h!]
\vspace{0.0cm}
\eject\centerline{\epsfig{file=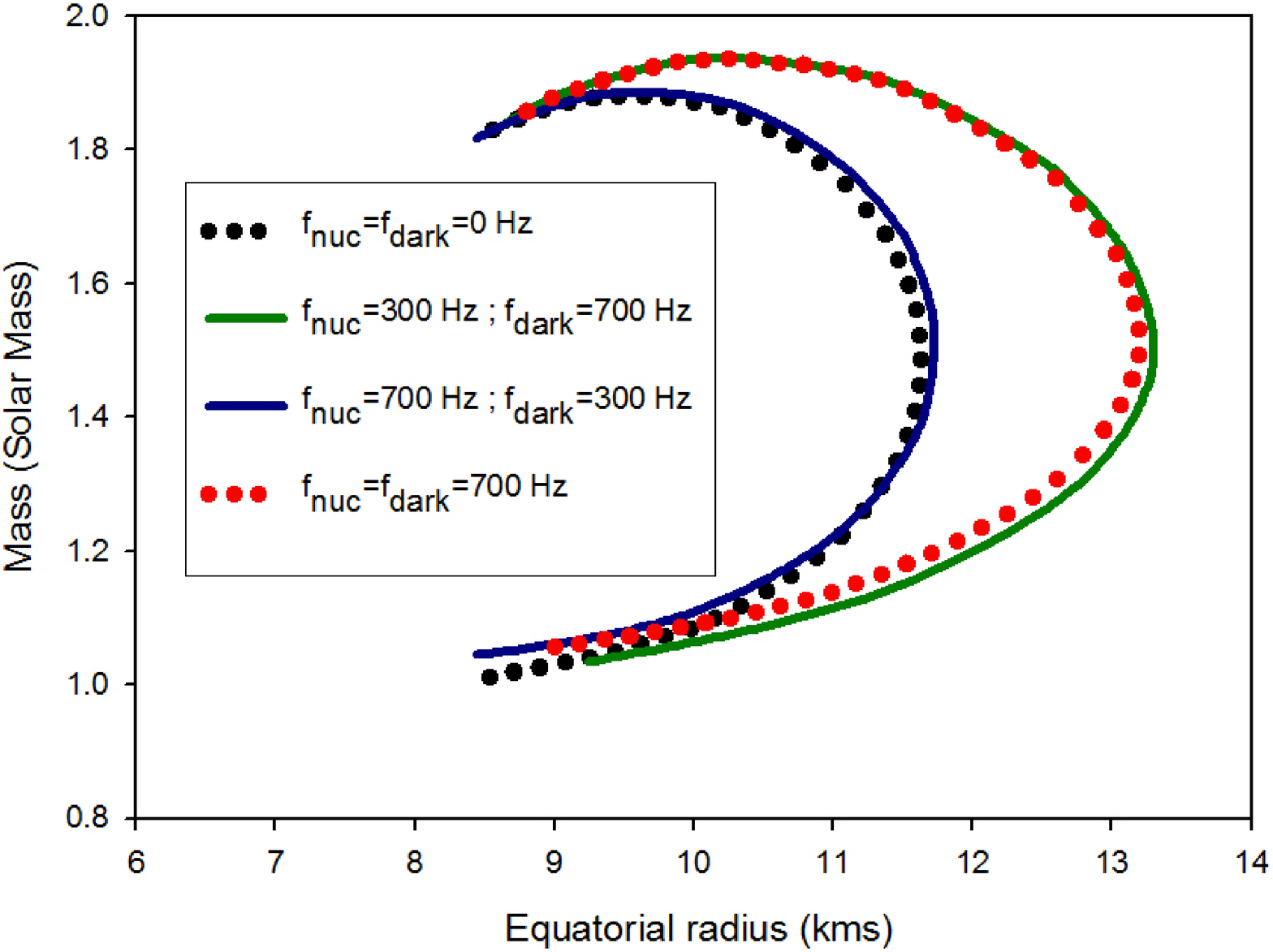,height=8cm,width=8cm}}
\caption
{Plots of total mass vs. equatorial radius of static, rigidly rotating and differentially rotating neutron stars mixed with interacting fermionic Asymmetric Dark Matter with fixed nuclear matter central enthalpy ($0.24c^2$) and varying dark matter central enthalpies.}
\label{fig7}
\vspace{0.0cm}
\end{figure}
\noindent 

\begin{figure}[h!]
\vspace{0.0cm}
\eject\centerline{\epsfig{file=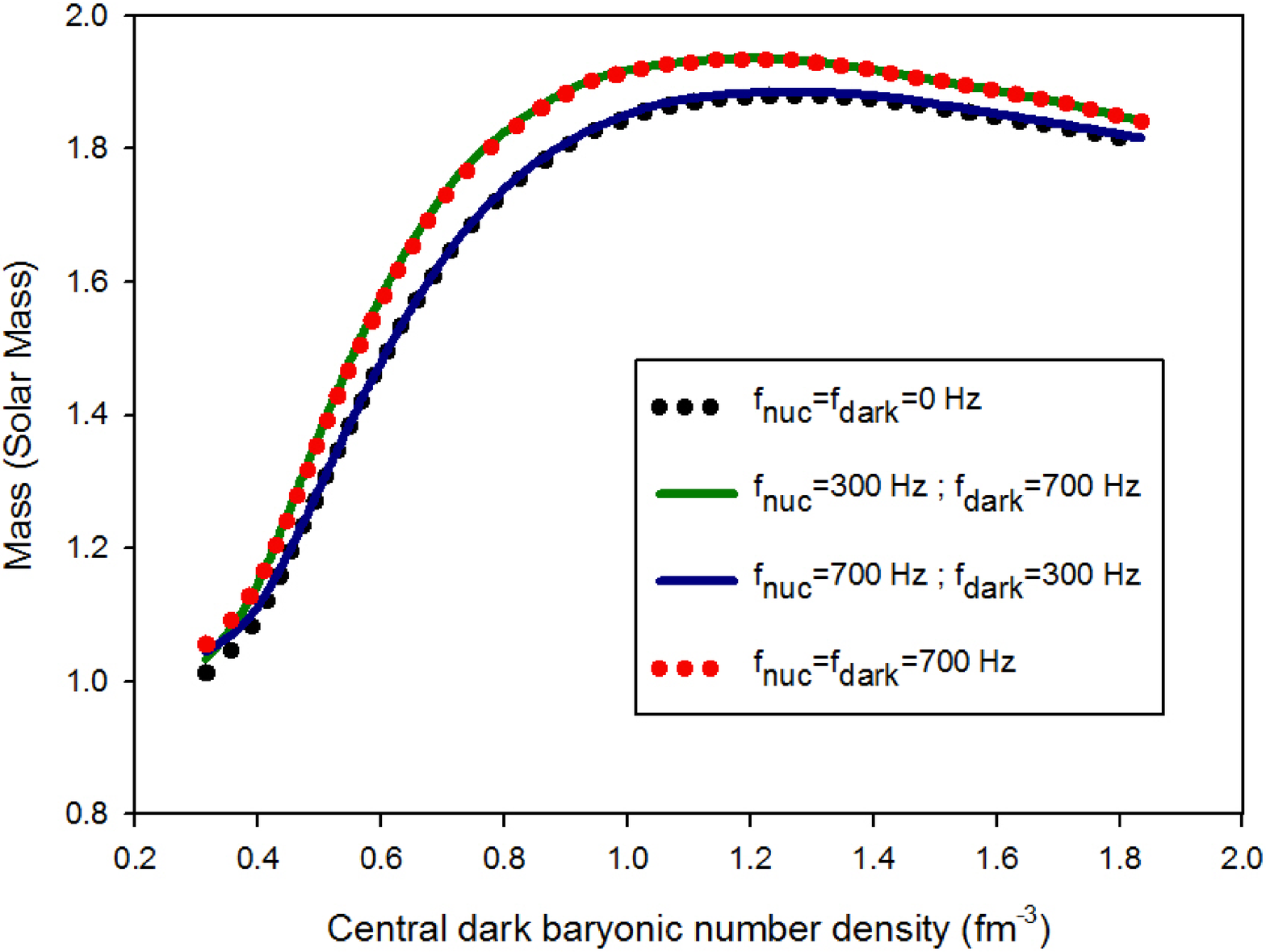,height=8cm,width=8cm}}
\caption
{Plots of total mass vs. central dark matter density of static, rigidly rotating and differentially rotating neutron stars mixed with self-interacting fermionic Asymmetric Dark Matter with fixed nuclear matter central enthalpy ($0.24c^2$) and varying dark matter central enthalpies.}
\label{fig8}
\vspace{0.0cm}
\end{figure}
\noindent

\noindent
\section{ Results and Discussions }
\label{section7}

    In Fig.-5 the plots of total mass vs. equatorial radius of static, rigidly and differentially rotating neutron stars mixed with fermionic self-interacting dark matter are shown for fixed dark matter central enthalpy $(0.24c^2)$ and varying nuclear matter central enthalpies. In Fig.-6 the corresponding plots of mass vs. central baryonic number density are shown. The maximum mass of the neutron star mixed with strongly self-interacting dark matter goes to $1.3640 M_\odot$ with a corresponding radius of $6.7523$ kms for the case of differential rotation (frequency of dark matter to be 300 Hz and that of nuclear matter to be 700 Hz) as shown in Fig.-5. From Fig.-6 we see that the corresponding central baryonic number density is $2.1060 fm^{-3}$. In this case, while the maximum gravitational mass is $1.3640 M_\odot$, the corresponding matter mass is $1.5024 M_\odot$ which constitutes of nuclear matter $1.4719 M_\odot$ and dark matter $0.0305 M_\odot$.  
		
		In Fig.-7 the plots of total mass vs. equatorial radius of static, rigidly and differentially rotating neutron stars mixed with fermionic self-interacting dark matter are shown for fixed nuclear matter central enthalpy $(0.24c^2)$ and varying dark matter central enthalpies. In Fig.-8 the corresponding plots of mass vs. central dark baryonic number density are shown. In this case the maximum mass goes to $1.9355 M_\odot$ with a corresponding radius of $10.3717$ kms for the case of differential rotation (frequency of dark matter to be 700 Hz and that of nuclear matter to be 300 Hz) as shown in Fig.-7. From Fig.-8 we see that the corresponding central dark baryonic number density is $1.1605 fm^{-3}$. For this case, while the maximum gravitational mass is $1.9355 M_\odot$, the corresponding matter mass is $2.1105 M_\odot$ which constitutes of nuclear matter $0.1179 M_\odot$ and dark matter $1.9926 M_\odot$.  
    
    It is seen that the polytropic indices $\gamma$ for nuclear and self-interacting dark matter EoSs are approximately equal, but the polytropic coefficient $K$ for dark matter is about 2.5 times larger than that of nuclear matter making dark matter EoS stiffer. Consequently, configurations of stars with varying dark matter central enthalpy with fixed nuclear matter central enthalpy are more massive than those obtained for the reverse case. 
    
    From Fig.-7 we see that the dark matter dominated neutron star behaves differently than the nuclear matter dominated one as shown in Fig.-5. In  Fig.-7, the plots of low mass neutron stars admixed with dark matter typically show characteristics similar to low mass self-bound strange stars. This is because of the very strong two-body repulsive interactions of dark matter which is dominant in the configuration of Fig.-7 which counteracts gravity effectively for low mass region and makes radius much smaller compared to pure neutron star of similar mass (vide Fig.-2). Thus, while the nuclear matter dominance induces gravitational binding, dark matter dominant low mass neutron star becomes gravitationally bound at much smaller radius.   
    
    The maximum mass for non-rotating dark matter stars goes to $3.0279 M_{\odot}$ with a radius of 16.2349 kms for particle mass $m_\chi=1$ GeV, and that for rotating stars it goes to $3.1460 M_{\odot}$ with a radius of 19.2173 kms. However, if one takes $m_\chi$ to be 0.5 GeV, then the maximum mass goes to $\sim 12.6 M_{\odot}$ using the relation Mass $\propto 1/m^2_\chi$ \cite{Na06}, thus mimicking stellar mass black holes.        	
   
\noindent    
\section{ Summary and Conclusions }
\label{section8}

    In this work we consider fermionic Asymmetric Dark Matter (ADM) particles of mass 1 GeV and the self-interaction mediator mass of 100 MeV (low mass implying strong interaction), mixed with rotating and non-rotating neutron stars. These ADM particles are non self-annihilating and behaves like ordinary free particles. We have shown that massive exotic neutron star with a strong two-body self-interacting fermionic dark matter is gravitationally stable with equal or unequal rotational frequencies of the two fluids. This provides an alternative scenario for the existence of $\sim 2 M_\odot$ neutron stars with `stiff' equations of states. 
    
    The mass-radius relations of pure hadronic stars mixed with self-interacting fermionic Asymmetric Dark Matter have been obtained using the LORENE code.  For the case of pure dark matter stars consisting of less massive dark particles we see that the maximum masses can be comparable to that of stellar mass black holes. For the case of hadronic stars mixed with dark matter, we considered three different configurations - static, rigid rotation and differential rotation of nuclear matter and dark matter fluids. From the results, we conclude that for the dark matter dominated configurations the masses are more, ${\it viz.}$ for the static case the maximum masses of these hybrid stars can reach upto $\sim 1.88 M_\odot$ with corresponding radii $\sim 9.5$ kms whereas in the rigid and differential rotational cases the maximum masses of these hybrid stars can reach upto $\sim 1.94 M_\odot$ with corresponding equatorial radii $\sim 10.4$ kms. 
    
    We also find that the dark matter dominated neutron star behaves differently than the nuclear matter dominated one that show characteristics similar to low mass self-bound strange stars. This is because of the very strong two-body repulsive interactions of dark matter which is dominant in the low mass region where it counteracts gravity effectively to make radius much smaller. Thus, while the nuclear matter dominance induces gravitational binding, dark matter dominant low mass neutron star becomes more compact. However, if the dark matter particle mass is small compared to the nucleon mass the maximum mass may well be above $2 M_\odot$, provided no phase transition from nuclear to quark matter occurs. 
    
    In the past, phase transition and the possible existence of a pion condensate or quark matter
inside compact stars have been studied extensively \cite{Ka81,Ka83,Ma05,Sh07,Zd13}. The phase transition from baryonic matter to quark matter is decided by the transition density. It is observed that deconfinement transition inside neutron star causes reduction in its mass. It would be interesting to find the effect of dark matter on the compact hybrid stars (baryonic matter with quark core), mixed with self-interacting fermionic asymmetric dark matter. Such effect cannot be predicted a priori without full calculations, and we leave it for future investigation.


\newpage    
\begin{thebibliography}{99}

\bibitem{Ut16} Y. Utsumi et al., {\it The Astrophysical Journal} {\bf 833}, 156 (2016).
	
\bibitem{Gi05} Gianfranco Bertone, Dan Hooper and Joseph Silk {\it Physics Reports} {\bf 405}, 279 (2005).

\bibitem{Ge84} George R. Blumenthal, S. M. Faber, Joel R. Primack and Martin J. Rees, {\it Nature} {\bf 311}, 517 (1984).

\bibitem{Ki12} S. King and A. Merle, {\it JCAP} {\bf 016}, 1208 (2012).

\bibitem{Ka14} Kathryn M. Zurek, {\it Physics Reports} {\bf 537}, 91 (2014).

\bibitem{Fr12} C. S. Frenk and S. D. M. White, {\it Ann. Phys.} {\bf 524}, 507 (2012).

\bibitem{Do08} Douglas Spolyar, Katherine Freese and Paolo Gondolo, {\it Phys. Rev. Lett.} {\bf 100}, 051101 (2008).

\bibitem{Al15} Alan H. Guth, Mark P. Hertzberg and C. Prescod-Weinstein, {\it Phys. Rev.} {\bf D 92}, 103513 (2015).

\bibitem{Li2012} X. Li, T. Harko and K. Cheng, {\it JCAP} {\bf 06}, 001 (2012).

\bibitem{St78} G. Steigman, C. Sarazin, H. Quintana and J. Faulkner, {\it Astron. J.} {\bf 06}, 1050 (1978).

\bibitem{Go89} I. Goldman and S. Nussinov, {\it Phys. Rev.} {\bf D 40}, 3221 (1989).

\bibitem{Be08} G. Bertone and M. Fairbairn, {\it Phys. Rev.} {\bf D 77}, 043515 (2008).

\bibitem{Fr10} M. T. Frandsen and S. Sarkar, {\it Phys. Rev. Lett.} {\bf 105}, 011301 (2010).

\bibitem{Ci11} P. Ciarcelluti and F. Sandin, {\it Phys. Lett.} {\bf B 695}, 19 (2011). 

\bibitem{Le11} S.-C. Leung, M.-C. Chu and L.-M. Lin, {\it Phys. Rev.} {\bf D 84},107301 (2011).

\bibitem{LWC12} X. Y. Li, F. Y. Wang and K. S. Cheng, {\it JCAP} {\bf 10}, 031 (2012).

\bibitem{La10} A. de Lavallaz and M. Fairbairn, {\it Phys. Rev.} {\bf D 81}, 123521 (2010).

\bibitem{Ko10} C. Kouvaris and P. Tinyakov, {\it Phys. Rev.} {\bf D 82}, 063531 (2010).

\bibitem{Ko11} C. Kouvaris and P. Tinyakov, {\it Phys. Rev.} {\bf D 83}, 083512 (2011).

\bibitem{Da15} David M. Jacobs, Glenn D. Starkman and Bryan W. Lynn, {\it MNRAS} {\bf 450}, 3418 (2015).

\bibitem{Ko08} C. Kouvaris, {\it Phys. Rev.} {\bf D 77}, 023006 (2008).

\bibitem{Mc10} M. McCullough and M. Fairbairn, {\it Phys. Rev.} {\bf D 81}, 083520 (2010).

\bibitem{Pe12} M. \'Angeles P\'erez-Garc\'ia and J. Silk, {\it Phys. Lett.} {\bf B 711}, 6 (2012).

\bibitem{Na06} Gaurav Narain, J\"urgen Schaffner-Bielich and Igor N. Mishustin {\it Phys. Rev.} {\bf D 74}, 063003 (2006).

\bibitem{Li12} Ang Li, Feng Huang and Ren-Xin Xu, {\it Astroparticle Physics} {\bf 37}, 70 (2012).

\bibitem{Qi14} Qian-Fei Xiang, Wei-Zhou Jiang, Dong-Rui Zhang and Rong-Yao Yang  {\it Phys. Rev.} {\bf C 89}, 025803 (2014).

\bibitem{Ha11} M. R. S. Hawkins, {\it Mon. Not. R. Astron. Soc.} {\bf 415}, 2744 (2011).

\bibitem{To15} Laura Tolos and J\"urgen Schaffner-Bielich, {\it Phys.Rev.} {\bf D 92}, 123002 (2015).

\bibitem{Go13} I. Goldman, R. N. Mohapatra, S. Nussinov, D. Rosenbaum, V. Teplitz  {\it Phys. Lett.} {\bf B 725}, 200 (2013).

\bibitem{Po15} Jason Pollack, David N. Spergel and Paul J. Steinhardt, {\it Astrophys. J.} {\bf 804}, 131 (2015).

\bibitem{Ho15} Yonit Hochberg, Eric Kuflik, Hitoshi Murayama, Tomer Volansky and Jay G. Wacker, {\it Phys. Rev. Lett.} {\bf 115}, 021301 (2015).

\bibitem{Ch16} Chian-Shu Chen, Guey-Lin Lin and Yen-Hsun Lin, {\it JCAP} {\bf 01}, 013 (2016).

\bibitem{Ja14} James M. Cline, Zuowei Liu, Guy D. Moore, and Wei Xue, {\it Phys. Rev.} {\bf D 90}, 015023 (2014).

\bibitem{Ki14} Kimberly K. Boddy, Jonathan L. Feng, Manoj Kaplinghat, and Tim M.P. Tait, {\it Phys. Rev.} {\bf D 89}, 115017 (2014).

\bibitem{Ro13} Miguel Rocha, Annika H. G. Peter, James S. Bullock, Manoj Kaplinghat, Shea Garrison-Kimmel, Jose O\~norbe and Leonidas A. Moustakas, {\it MNRAS} {\bf 430}, 81 (2013).

\bibitem{Er10} Eric Gourgoulhon {\it arXiv:1003.5015v2} (2011).

\bibitem{La62} A. M. Lane, {\it Nucl. Phys.} {\bf 35}, 676 (1962). 

\bibitem{Sa83} G. R. Satchler, {\it Int. series of monographs on Physics}, Oxford University Press, {\it Direct Nuclear reactions}, 470 (1983).

\bibitem{BCS08} D. N. Basu, P. Roy Chowdhury and C. Samanta, {\it Nucl. Phys.} {\bf A 811}, 140 (2008).

\bibitem{Be77} G. Bertsch, J. Borysowicz, H. McManus, W. G. Love, {\it Nucl. Phys.} {\bf A 284}, 399 (1977).

\bibitem{Sa79} G. R. Satchler and W. G. Love, {\it Phys. Reports} {\bf 55}, 183 (1979). 

\bibitem{Sa89} C. Samanta, D. Bandyopadhyay and J. N. De, {\it Phys. Lett.} {\bf B 217}, 381 (1989). 

\bibitem{CB06} P. Roy Chowdhury and D. N. Basu, {\it Acta Phys. Pol.} {\bf B 37},1833 (2006).

\bibitem{Au03} G. Audi, A. H. Wapstra and C. Thibault, {\it Nucl. Phys.} {\bf A 729}, 337 (2003).

\bibitem{Lu03} D. Lunney, J. M. Pearson and C. Thibault, {\it Rev. Mod. Phys.} {\bf 75}, 1021 (2003).

\bibitem{Ro06} G. Royer and C. Gautier, {\it Phys. Rev.} {\bf C 73}, 067302 (2006).

\bibitem{Li07} T. Li, U. Garg, Y. Liu et al., {\it Phys. Rev. Lett.} {\bf 99}, 162503 (2007).

\bibitem{FMT49} R. P. Feynman, N. Metropolis and E. Teller, {\it Phys. Rev.} {\bf 75}, 1561 (1949).

\bibitem{BPS71} G. Baym, C. J. Pethick and P. Sutherland, {\it Astrophys. J.} {\bf 170}, 299 (1971).

\bibitem{BBP71} G. Baym, H. A. Bethe and C. J. Pethick, {\it Nucl. Phys.} {\bf A 175}, 225 (1971).

\bibitem{At17} D. Atta, S. Mukhopadhyay and D. N. Basu, {\it Indian J. Phys.}, {\bf 91}, 235 (2017).

\bibitem{At14} Debasis Atta and D. N. Basu, {\it Phys. Rev.} {\bf C 90}, 035802 (2014). 

\bibitem{Se14} W. M. Seif and D. N. Basu, {\it Phys. Rev.} {\bf C 89}, 028801 (2014).

\bibitem{Ch09} P. Roy Chowdhury, D. N. Basu and C. Samanta, {\it Phys. Rev.} {\bf C 80}, 011305(R) (2009).

\bibitem{Ba09} D. N. Basu, P. Roy Chowdhury and C. Samanta, {\it Phys. Rev.} {\bf C 80}, 057304 (2009).

\bibitem{St} Stuart L. Shapiro and Saul A. Teukolsky {\it Black Holes, White Dwarfs and Neutron Stars: The Physics of Compact Objects }.

\bibitem{Sa09} F. Sandin and P. Ciarcelluti, {\it Astroparticle Physics} {\bf 32}, 278 (2009).

\bibitem{Ka81} B. K\"ampfer, {\it J. Phys.} {\bf A 14}, L471 ( 1981).

\bibitem{Ka83} B. K\"ampfer, {\it J. Phys.} {\bf G 9}, 1487 (1983).

\bibitem{Ma05} Jean Macher and J\"urgen Schaffner-Beilich, {\it Eur. J. Phys.} {\bf 26}, 341 (2005).

\bibitem{Sh07} B. K. Sharma, P. K. Panda, and S. K. Patra, {\it Phys. Rev.} {\bf C 75}, 035808 (2007).

\bibitem{Zd13} J. L. Zdunik, P. Haensel, {\it A$\&$A} {\bf 551}, A61 (2013).

\end{thebibliography}
\end{document}